\documentclass{elsart}
\usepackage{natbib}
\usepackage{epsfig}
\usepackage{amssymb}

\begin{document}

\newcommand{\nat}{Nat}    
\newcommand{\apj}{ApJ}    
\newcommand{\aap}{A\&A}   
\newcommand{\mnras}{MNRAS}
\newcommand{\newa}{NewA}  
\newcommand{\pasj}{PASJ}  
\newcommand{\araa}{ARA\&A} 
\newcommand{\cal}{}
\newcommand{\txi}{\tilde{\xi}}
\newcommand{\bxi}{{\bf \xi}}
\newcommand{\bmath}{\bf}
\newcommand{\bGamma}{{\bf \Gamma}}
\newcommand{\tw}{\tilde{w}}
\newcommand{\balpha}{{\bf \alpha}}
\newcommand{\bX}{{\bf X}}
\newcommand{\xp}{{\bf\hat x'}}
\newcommand{\yp}{{\bf\hat y'}}
\newcommand{\xh}{{\bf\hat x}}
\newcommand{\yh}{{\bf\hat y}}
\newcommand{\bnodes}{{\bf nodes}}
\newcommand{\beq}{\begin{equation}}
\newcommand{\eeq}{\end{equation}}

\begin{frontmatter}

%

\title{Fast n-point correlation functions and three-point lensing application}
\author[CITA,petfat]{Lucy Liuxuan Zhang}
\author[CITA,pen]{Ue-Li Pen}
\address[CITA]{Canadian Institute for Theoretical Astrophysics,\\
		University of Toronto, M5S 3H8, Canada}
\address[petfat]{lxzhang@cita.utoronto.ca}
\address[pen]{pen@cita.utoronto.ca}

\date{version 15 Feb 2003}
\pubyear{2004}
\label{firstpage}

\begin{abstract}
We present a new algorithm to rapidly compute the two-point (2PCF),
three-point (3PCF) and $n$-point ($n$-PCF) correlation functions in
roughly ${\cal O}(N\log N)$ time for $N$ particles,
instead of ${\cal O}(N^n)$ as required by brute force approaches.
The algorithm enables an estimate of the full 3PCF for as many
as $10^6$ galaxies.
This technique exploits node-to-node correlations of a recursive
bisectional binary tree.  A balanced tree construction
minimizes the depth of the tree and the worst case error at each node.
The algorithm presented in this paper can be applied to problems with 
arbitrary geometry.

We describe the detailed implementation to compute the two point
function and all eight components of the 3PCF for a two-component
field, with attention to shear fields generated by
gravitational lensing.
We also generalize the algorithm to compute the $n$-point correlation
function for a scalar field in $k$ dimensions where $n$
and $k$ are arbitrary positive integers.
\end{abstract}
\end{frontmatter}

\section{Introduction}

The correlation functions are important tools for computation
and analysis in many areas of astrophysics.
They are introduced into modern cosmology by people such as
\citet{1969PASJ...21..221T} and \citet{1973ApJ...185..413P}.
Some prominent applications of the correlation functions
can be found in areas such as galaxy clustering, CMB, and weak lensing.
For a scalar field
$\rho(\bX)$, the two-point correlation function (2PCF) $\xi_2$ and
the three-point correlation function (3PCF) $\xi_3$ are defined as
\beq
  \xi_2(s)=\langle\rho(\bX_1)\rho(\bX_2)\rangle
\eeq
and
\beq
  \xi_3(s_1,s_2,s_3)=\langle\rho(\bX_1)\rho(\bX_2)\rho(\bX_3)\rangle
\eeq
where $s$ is the distance between $\bX_1$ and $\bX_2$, and $s_i$
represents the distance between $\bX_j$ and $\bX_k$ with
$i\neq j\neq k$.

The $n$-point correlation function ($n$-PCF)
of a scalar field $\rho(\bX)$ is a function of all the possible
arrangements of $n$ points chosen from the system.  As a function of
configurations, the value of the $n$-point function is the expectation
value of the product of the field quantity $\rho$,
sampled at a set of $n$ points
conforming to a specific configuration.
The one-point correlation function (1PCF) is simply
the weighted average of
$\rho$ over all data points; in the case of the 2PCF, the
spacial configuration is characterized by the separation
between two points.  From the definition of the ensemble
average, the $n$-point correlation function for a system of $N$ points
can be computed by looping through all combinations of $n$ points,
determining the configuration category for each combination,
and accumulating necessary statistics,
such as the sum of products of the weighted $\rho$'s and that
of the weights, under the appropriate configuration category.
Outside of the loop, we divide the sum of the products of the weighted
$\rho$'s by the sum of the products of their weights for each
configuration to obtain the expectation value of the product of
$\rho$'s at any $n$ points as a function of configurations.
This brute force approach amounts to a loop through
$N(N-1)...(N-n+1)$ combinations, which is
intrinsically an ${\cal O}(N^n)$ computation.
For $N$ of astronomical interest, such signifcant cost
makes it impossible to compute by brute force the full
correlation function of order $n$ greater than two.

In fact, even for n=2, more efficient algorithms are often necessary
and have been developed.
Early applications \citep{2000Natur.405..143W} of the two point
function in weak gravitational lensing summed pairs,
which is an ${\cal O}(N^2)$ algorithm.  Since an auto-correlation is a
convolution, several procedures exist to reduce it to
${\cal O}(N\log N)$ using either the well-known fast
Fourier transforms or a tree summation \citep{1986Natur.324..446B}.
See \citet{2003ApJ...592..664P} for an example of a 2PCF
algorithm in 2D using FFT.
In the astrophysical literature, algorithms to compute correlations of
arbitrary order have been presented by \citet{2001misk.conf...71M}.
These are reported to scale as ${\cal O}(N^{3/2})$ for the 2PCF
computation.

Recent progress has been made in effort to speed up the computation
of the three-point correlation function.
After this work was completed, we became aware of an independent
implementation of a very similar algorithm for the 3PCF
by \citet{2003astro.ph..7393J} based on an
${\cal O}(N\log N)$ implementation of the \citet{2001misk.conf...71M} 
algorithm.
Recently, an unpublished ${\cal O}(N)$ algorithm for computing
the $n$-point correlation function is also reported to exit
\citep{2004astro.ph..1121G}.  All of these algorithms are based on
the idea of kd-trees which is first proposed by
\citep{DBLP:journals/cacm/Bentley75}.

In this paper, we will describe a new, general and fast algorithm
for the $n$-PCF computation in arbitrary dimensions, with focus on the
implementation of correlation functions
in 2D weak lensing data analyses.
As mentioned in \citet{2003A&A...397..809S},
the two-point correlation function has been a popular method to
analyze lensing data because it can be easily
observed and cheaply computed; in addition, all second-order
statistical measures can be derived in terms of the
two-point function.
On the other hand, \citet{2003ApJ...592..664P} used
the three-point correlation function to measure the
skewness of the shear which helps break the
${\Omega\--\sigma_8}$ degeneracy.  As expected, the 3PCF is
computationally much more complex and expensive.
Most detections of the 3PCF or higher moments
merely take into account sparse subsets of the possible configurations,
such as the equilateral triangles \citep{2002PhRvL..88x1302S}.
Even the computation of a partial 3PCF required the use of
a supercomputer \citep{2001MNRAS.325..463S}.
The challenge of computing the {\it full} 3PCF is also illustrated
by \citet{2002MNRAS.335..432V}, who used only two different
triangle configurations to estimate the bispectrum, significantly
reducing the number of triangles from $75\ 792^3\sim 4\times 10^{14}$
to $80\times 10^6$.
\citet{2003astro.ph.10831E} estimated the n-PCF in ${\cal O}(N^n)$
time with a small prefactor for Monte Carlo analysis.  They also
pointed out that higher order $n$-point correlation functions are
notorious for being computationally expensive.
Besides its speed, our tree method of computing the 3PCF is
advantageous to the fourier space computation in that it is
free of geometric restrictions.  The tree structure allows
easy extension to computing correlation functions
of arbitrary order in higher dimensional space.

Fortunately, the new algorithm developed here makes it tractable
to handle rapidly increasing cosmological data sets.
Now using a single-processor mechine,
we are able to compute the full shear 3PCF for $10^{6}$ 
galaxies.
Besides its speed, it is superior to the existing
3PCF FFT algorithm \citep{2003ApJ...592..664P} in its freedom from
geometric restrictions.

This paper consists of six sections.
In section \ref{sec:buildtree},
we describe in detail the construction of the
recursive bisectional binary tree which is the basis for
rapid computation of the $n$-PCF.  In section \ref{sec:2PCF}, we show
how to compute the 2PCF using the tree by first discussing
the process of 'subdivision' and its dependence on the
'critical open angle'\footnote{The term 'open angle'
is coined by \citet{1986Natur.324..446B}.},
then reviewing the accumulation of two-point
statistics and, for a spin-2 field, our choice of coordinates for a
given pair of nodes.
In section \ref{sec:3PCF}, we conveniently extend the method for
computing the 2PCF to accommodate the 3PCF.
The algorithm is further generalized in section \ref{sec:nPCF}
to compute correlation functions of arbitrary order
in a $k$-dimensional space.
Section \ref{sec:speed} is devoted to address the speed and
efficiency of the overall performance, as well as the accuracy, with
empirical performance results from 3PCF computations.
Finally, we review all the procedures taken to compute the correlation
functions, and conclude with potential applications of the method in
section \ref{sec:conclusion}.

\section{Build Tree in 2D}
\label{sec:buildtree}

We use the monopole bisectional binary tree described in \citet
{2003MNRAS.346..619P} and node-to-node interactions to compute
the $n$-PCF.

The tree needs be constructed only once, and remains useful
thereafter for computing correlation functions of arbitrary order.
For this reason, we should invest
sufficient computing resources into this step to minimize the
worst case error arising from the utilization of the tree method. 
In order to achieve this,
we exploit a monopole tree decomposition.

One starts by defining a 'root' node
which contains all particles.
For nodes where only one particle is present,
we simply store the information about that particle with the node.
We never subdivide a single-particle node.

\begin{figure}
\vskip 3.3 truein
\includegraphics{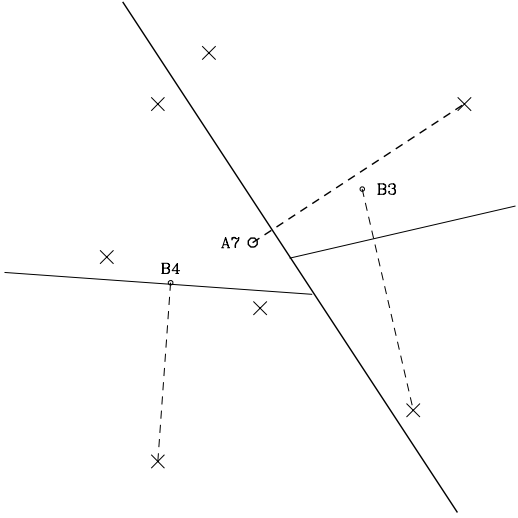}
\caption{For a set of seven points, we first find the weighted
average position; in this plot, it is simply the centre of mass,
labeled by A7,
because we choose equal weight for each point.
We subdivide the root node by cutting through a projected median point
perpendicular to the line connecting the weighted centre of mass
and the furthest particle from it, that is, the dashed line emitting
from A7.  We then recursively subdivide each of the daughter nodes
in the same manner. The character-number labels
in this figure indicate the placement of the nodes, whose centres
of mass coincide with the small circles, in the tree hierarchy,
as shown in figure \ref{fig:buildtree2}.}
\label{fig:buildtree1}
\end{figure}

For any node containing more than one particle,
the longest node extent is defined as the line joining
the weighted centre of mass (COM) and the particle furthest from
the COM; the length of the longest node extent is called
the size of the node.
We orthogonally project all the particles onto the longest node
extent, and choose the 'projected median point' to be a point on the line
which devides the longest node extent evenly in the following sense.
In the case where the number of particles in the node is even, the
'projected median point' devides the longest node extent into two rays
which contain an equal number of projected particles; otherwise,
the particles in one ray would outnumber those in the other ray by
exactly one.
When it is necessary to subdivide the node, we spatially
separate the node into two subnodes by a cut perpendicular to
the longest node extent and through the 'projected median point'.
This method of partitioning space is demonstrated for a sample
of seven points in figure \ref{fig:buildtree1}.
When walking the tree, the size of a node is a crucial piece
of information for the 'open angle'
test to be discussed in section \ref{sec:theta}.

\begin{figure}
\vskip 3.3 truein
\includegraphics{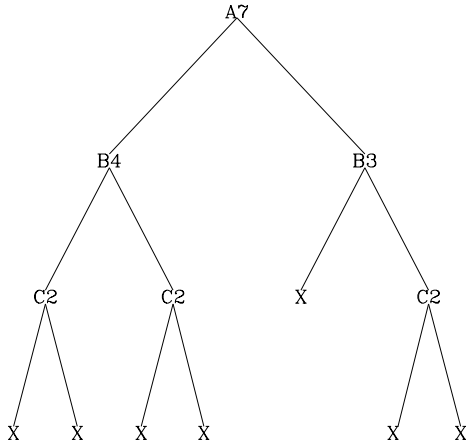}
\caption{The figure shows the population of a bisectional binary
tree constructed for seven particles.  The number at each node
indicates the number of particles occupying that node.
We draw our tree upside down following the convention of computer
scientists.
In our discussions, A is called the 'root' node and X the leaves;
thus, descent means moving from the root toward the leaves.}
\label{fig:buildtree2}
\end{figure}

For a data set containing $N$ particles where $N$ is an integer power
of $2$, we always find an exactly equal number of particles in all
nodes situating at the same level of the tree hierarchy;
otherwise, it is possible for a minority (ie. less or equal to half)
of the nodes to contain one fewer particle than the other nodes
at the same level.  The bisectional binary tree constructed
for seven particles is populated as shown
in figure \ref{fig:buildtree2}.
The resulting binary tree is balanced, and the depth of
the tree is pre-determined to be $1+\log_2N$ rounded up
to the next integer.
This tree is fast to build.
At each level of the tree,
the sorting and bisection only costs ${\cal O}(N \log N)$ time
for $N$ particles.  Since there are $\sim\log_2N$ levels
in the tree, the total cost of building the tree is
${\cal O}(N (\log N)^2)$.

We keep track of the position of a node in the hierarchical tree
structure by using an index lookup table.
The index lookup table is a $(2N-1)\times2$ array,
intended to contain the indices of the two subnodes for
each of the $2N-1$ non-empty nodes.
A node is either a leaf, when it contains a single particle,
or a compound node, if it is composed of two distinct subnodes.
We reserve the first $N$ rows of the table for the $N$ leaves,
numbered in the order of input,
and assign to their entries the value of their own index.
The next $N-1$ rows carry indices of the subnodes belonging to
the remaining $N-1$ compound nodes which are numbered dynamically
from the 'root' down during the process of tree building.
In general, the entries in the $i^{th}$ row of the table
hold the indices of the two daughter nodes of the $i^{th}$ node.
Near the bottom of the tree where a node contains a leaf
as one of its subnodes, the corresponding entry in the index
lookup table yields an index smaller than $N$,
thus pointing back to the first $N$ rows of the table.
When walking the tree, the relation that the index of a node is
smaller than $N$ is often interpreted as a terminating signal for
subdivisions which we will discuss further in section \ref{sec:sub2}.

The weight of a compound node is the sum of the
weights of its member particles.
For the purpose of weak lensing data analysis, we choose the
weight of a particle, ie. a galaxy, to be the inverse noise squared.
For a single-particle node, the size of the node is the same
as that of the particle.  The particle size only has
to be small enough so that it does not interfere with the
'open angle' test, and nonzero for numerical stability.
Hence, we treat the particle roughly like a point,
with a small but nonzero extent of $10^{-5} \times 0.2''$.
This size is much smaller than $2''$ which is the minimum
separation resolved by the survey, or in other words,
the smallest distance between any non-degenerate pair
galaxies.

We summarize the useful information to be stored in each indexed
node as follows:
\begin{description}
  \item the weighted centre of mass
  \item the size of the node
  \item the weight of the node
  \item other weighted average quantities of interest.
\end{description}

\section{Two-Point Correlation Function}
\label{sec:2PCF}

The two-point correlation function has a wide range of applications,
including that in weak gravitational lensing data analysis.
It can be computed cheaply, and is relatively
easy to obtain from lensing surveys.
Here, we discuss a node-node 2PCF construction.  This approach
is different from the 2PCF computation by \citet{2003ApJ...592..664P}
which is ${\cal O}(N\log N)$ and particle-node based.
This node-node method costs ${\cal O}(N\theta_c^{-2})$ once the tree
has been built (see section \ref{subsec:estimate_speed} for derivation).
For a flat survey geometry,
the two-point correlation function can be computed rapidly using
Fourier Transforms with computing time ${\cal O}(N \log N)$.
However, since the geometry of the sky is non-Euclidean,
we cannot apply the FFT method to the problem (except as noted
by \citet{2003NewA....8..581P}).
Then the geometry-independent 2PCF algorithm presented
in this section, whose computing cost
has a similar $N$-dependence as the FFT method though with a larger
prefactor, becomes a favourable alternative.
This 2PCF algorithm makes use of the bisectional binary tree
constructed in section \ref{sec:buildtree}.

\subsection{\bf Subdivision for 2PCF}
\label{sec:sub2}
In the mainstream subdivision,
we decide which pairs of nodes should be correlated.
Once a pair of nodes is put into this category,
we then determine whether they can be directly used to
compute the two-point correlation.  In the case where
the nodes are sufficiently far apart from each other compared to
their size, we directly correlate between them; otherwise,
we further subdivide these nodes, and
compare their sub\-nodes {\it instead}.
As noted by \citet{1989ApJS...70..389B}, hierarchical methods are
based on the observation that, when calculating the interaction
between particles, it makes sense to ignore the detailed internal
structure of distant groups of particles.  More radical than using
particle-group interactions, we adopt group-group interations for
distant groups in place of many particle-particle interactions between
the groups.  Utilizing the tree-shaped data structure, the computation
of the $n$-PCF, which boils down to a great number of particle
interactions, can be achieved with very significant computational savings.

When we come to a node, the situation is analogous to the
dilemma where two people come
to a three-way intersection, and must keep going forward.
If we label the streets that are in front of them $s_1$
and $s_2$, and themselves $q_1$ and $q_2$,
all possibilities of subsequent actions can be summarized as
\begin{description}
\item $q_1,q_2 \rightarrow s_1$
\item $q_1,q_2 \rightarrow s_2$
\item $q_1 \rightarrow s_1, q_2 \rightarrow s_2$
\item $q_1 \rightarrow s_2, q_2 \rightarrow s_1$.
\end{description}
If the persons are indistinguishable, the last two possibilities
appear to be the same, then the number of different combinations is
reduced from four to three.
In our language, the nodes are equivalent to the persons, and
the paths through which the nodes are
processed are analogous to the streets.
Both the mainstream subdivision and sbusequent processings
are independent of the ordering of arguments, which implies that
the correlation between each pair would be undesirably weighted
twice if the nodes were considered distinct.
Hence, for our purpose, any two nodes are considered
indistinguishable with respect to the subdivision process.

All possibilities can be accounted for by using a 
recursive subroutine called SUBDIVIDE2 which takes two
input arguments indicating the indices of the nodes to be compared.
Let us name the two input nodes $p_1$ and $p_2$,
and their subnodes $p_1^j$ and $p_2^j$ respectively,
where $j \in \{1,2\}$ labels the two subnodes.
We shall refer to the indices of these nodes as $k_1$ and $k_2$,
and those of the subnodes $k_1^j$ and $k_2^j$.
We initiate the subdivision process by calling SUBDIVIDE2($N+1,N+1$)
which correlates all particles within the 'root' node with index $N+1$.
Then the recursive subdivisions take over and complete the rest of
the necessary subdivisions for us.

\subsubsection{\bf Mainstream subdivision}
In what we call the mainstream subdivision, we correlate {\it between}
and {\it within} nodes at every level of the tree hierarchy
to ensure full coverage of space and separation lengths.

{\it Case1} $\Leftrightarrow$ SUBDIVIDE2($k_1,k_2 | k_1=k_2$):
If we want to correlate within a node, such as the 'root' node,
we call SUBDIVIDE2 with $k_1=k_2$, each of which equals the
index of the node.
In this case, we only have three ways of choosing two points from
its subnodes, all of which should be explored:
\begin{description}
\item Choose 2 points from $p_1^1$ if $k_1^1 > N$\\
   $\Leftrightarrow$ SUBDIVIDE2($k_1^1,k_1^1$),
\item Choose 2 points from $p_1^2$ if $k_1^2 > N$\\
   $\Leftrightarrow$ SUBDIVIDE2($k_1^2,k_1^2$),
\item Choose 1 point from $p_1^1$ and 1 from $p_1^2$\\
   $\Leftrightarrow$ SUBDIVIDE2($k_1^1,k_1^2$).
\end{description}

Note that $k_1=k_2$ are always greater than N by construction.
The first two possibilities lead back to {\it Case1} at the
next level.
However, the third option pipes the subnodes
to the further subdivision process since it explicitly specifies
two distinct nodes to correlate {\it between}.

\subsubsection{\bf Further subdivision}
\label{sec:theta}
{\it Case2} $\Leftrightarrow$ SUBDIVIDE2($k_1,k_2 | k_1\neq k_2$):
When the two arguments of SUBDIVIDE2 are different,
implying $p_1 \neq p_2$,
we are admitted into the further subdivision process,
which depends on the 'open angle' criterion.
In this step, we want to decide whether this pair of nodes
is suitable to be used in computing the two-point correlation.
The 'critical open angle' ${\theta_c}$ is an accuracy parameter that 
controls the further subdivision process.
We define, for example, the open angle of node 1 relative to node 2 as
the ratio of the size of node 1, $r_1$, to the distance between
the nodes, $d$.  If this ratio, $\frac{r_1}{d}$, is greater than
the critical angle $\theta_c$,
then node 1 is considered too large to be used as
a unit, and we subdivide it into its two subnodes.
We refer to this test as the 'open angle criterion'.
As described above, the 'critical open angle', $\theta_c$
is the {\it maximal tolerable} ratio of the size of one node to
its distance from the other node,
to be satisfied by any pair of nodes being correlated.
This is similar to the
concept of truncation error in numerical simulations.  Geometrically,
$\theta_c$ represents the critical linear angle as seen from
the particle we are testing\footnote{The concept of the 'open angle'
and some advantages of the tree scheme in terms of solving the
gravitational $N$-body problem have been thoroughly explained by
\citet{1989ApJS...70..389B}.}.
Given a pair of nodes, $p_1$ and $p_2$, let $r_i$ denote the
size of $p_i$ and $d$ the separation distance between $p_1$
and $p_2$.
If both nodes meet the 'open angle' criterion, we perform
the node-to-node two-point correlation (2PC) on them;
otherwise, we subdivide the nodes into subnodes until each node
in a pair fromed by the subnodes satisfies the criterion.
Each round the further subdivision is invoked, we only check
whether the {\it first} node satisfies the
'open angle' criterion.  However, both nodes should be tested
before 2PC is performed on them.
This is achieved by interchanging the order in which the pair of nodes
is passed to SUBDIVIDE2 once the first node has met the criterion.
To prevent infinite interchanges of the arguments and repeated
calls of SUBDIVIDE2 even when both nodes have satisfied the criterion,
we introduce a counter $c$, which is intially set to zero, to
keep track of the number of times the criterion has been fulfilled
by a node from the same pair.  This is put into code format as
follows with the additional argument $c$. \\
\\
In SUBDIVIDE2($k_1,k_2,c$) with $k_1\ne k_2$:\\
If $\frac{r_1}{d}\leq\theta_c$ or $k_1\le N$, we have
\begin{verbatim}
  If (c=1) then
    call 2PC(k1,k2)
    return
  Else
    call SUBDIVIDE2(k2,k1,c+1)
  Endif.
\end{verbatim}

If $\frac{r_1}{d}>\theta_c$ and if $k_1 > N$,
we subdivide $p_1$, and correlate between
all pairs formed by $p_2$ and a subnode of $p_1$ by carrying out the
following actions.
\begin{description}
\item Choose 1 points from $p_1^1$ and 1 from $p_2$\\
   $\Leftrightarrow$ SUBDIVIDE2($k_1^1,k_2,c=0$),
\item Choose 1 points from $p_1^2$ and 1 from $p_2$\\
   $\Leftrightarrow$ SUBDIVIDE2($k_1^2,k_2,c=0$).
\end{description}

We set the counter to zero when a new pair is to be considered.
During these accuracy-dependent further subdivisions,
we do not compare subnodes belonging to the same parent node
as this would interfere with the duty of the mainstream subdivision.
All pairs of nodes satisfying the 'open angle' criterion are
sent to the node-to-node two-point correlation (2PC)
discussed in the next section.

The combined subdivision process is designed such that,
when $\theta_c$ is assigned a sufficiently small value,
all pairs of individual particles are taken into account,
as in the brute force approach, to yield the exact 2PCF.

\subsection{\bf Node-to-node two-point correlation in 2D}
\label{sec:nn2}

In the subroutine 2PC, given a pair of distinct nodes,
we add the product of their weighted field quantities
and that of their weights to the respective accumulated
sums binned according to the pair separation.
The final 2PCF is obtained by taking the quotient of these two sums,
and should be independent of the field orientation.
In weak lensing analysis, both scalar fields, eg. $\kappa(\bX)$,
and spin-2 fields, eg. $\gamma(\bX)$,
are frequently encountered.
The 2PC for a scalar field is streight-forward.
It requires a unique identification of pair configurations,
regardless of its orientation in the physical space, with
a suitable binning.
The 2PC for a spin-2 field, however, requires additional treatment.
For a spin-2 field, a coordinate change is necessary
to eliminate the effect due to difference in the pair orientation,
and a choice should be made about what independent components
of the 2PCF are to be stored.

\subsubsection{\bf 2PC for a 2D scalar field}

In weak gravitational lensing,
the scalar field $\kappa$, which is the projected mass density,
is an important observable quantity.
The computation of the two-point correlation function for
such 2D scalar fields is thus a basic but essential task.
Here, using the $\kappa$ field as an example,
we illustrate an orientation independent way to map pairs
to the corresponding configuration space with a logarithmical
binning.

We use a simulated projected mass density map consisting of
up to a million galaxies.
The quantities used for the 2PC computation are
the position ($x$, $y$), $\kappa$, and the noise
for each galaxy.
When analyzing weak lensing data,
the weight of each individual galaxy is taken to be the inverse
noise squared.  Recall that in the tree construction, we sum up
the weights of all galaxies in the node, and store
the sum as the weight of the node.  Aside from the weight and
the size of a node, whose definitions are given in
section \ref{sec:buildtree}, other
quantities associated with the node are weighted averages. 

The configuration space for the 2PCF in 2D is one-dimensional,
and can be parameterized by the distance between the pair.
For computational convenience, we logarithmically bin the separation
between the pair of points.
The final two point function is a one-dimensional scalar function
of logarithmically binned intervals of separation distances,
given by

\beq
\xi_2=\frac{\txi_2}{\tw_2}
\eeq

where, for all pairs of nodes, $A$ and $B$, belonging to a given
configuration interval $\eta$,

\beq
\txi_2(\eta)=\sum{\bar\kappa(A)\bar\kappa(B)}
\label{eq:kappa}
\eeq
and
\beq
\tw_2(\eta)=\sum w(A)w(B).
\eeq
In equation \ref{eq:kappa}, $\bar\kappa$ is the weighted sum
of the $\kappa$ values which belong to the particles in the node.
This weighted sum is obtained by multiplying the weighted average
of the quantity $\kappa$ at the node by the total weight of the node.

\subsubsection{\bf 2PC for a 2D spin-2 field}

In this section, we show how two-point correlation statistics
can be accumulated when applied specifically to a spin-2
lensing shear map.
We use a simulated shear map consisting of up to a million galaxies.
The quantities obtained from the simulation are the position ($x$, $y$),
the shear ($\gamma_1$, $\gamma_2$), and the noise for each galaxy.
We continue to use the parameterization and the logarithmical binning
described in the previous section.
However, for a spin-2 field, we also need to perform 
a coordinate transformation to ensure the rotational invariance
of the 2PCF.

The original coordinate system is chosen such that all galaxies
have positive $x$ and $y$ as components of their position
when the data is stored.
We are free to do so since the correlation function is invariant
in translations of the coordinate system.
The shears are conventionally recorded as ($\gamma_1$, $\gamma_2$)
in Cartesian coordinates, like that in our simulation.
However it can also be expressed in polar coordinates as
($\gamma$, $\phi$).
The conversions between these two bases are

\beq
\gamma_1=\gamma\cos{2\phi},
\label{eq:gamma1}
\eeq
\beq
\gamma_2=\gamma\sin{2\phi}
\label{eq:gamma2}
\eeq

and

\beq
\gamma=\sqrt{\gamma_1^2+\gamma_2^2},
\label{eq:gamma3}
\eeq
\beq
\tan(2\phi)=\frac{\gamma_2}{\gamma_1}.
\label{eq:gamma4}
\eeq

Here, we correlate the shear components in terms of a special
angular coordinate.
For any two points, $A$ and $B$, we first convert each of their shears
given as ($\gamma_1$, $\gamma_2$) into 
the polar components ($\gamma$, $\phi$).
While keeping $\gamma$ fixed at each point,
we subtract from their original $\phi$ the angle $\beta\in\{0,\pi\}$
between $\bf\hat{x}$ and the line connecting the two points.
We then convert the rotated shears back into Cartesian coordinates,
and denote them as ($\gamma_1'$, $\gamma_2'$).  These rotated
shears are used to compute the two-point correlation function.

The 2PCF of a spin-2 field in 2D has two {\it independent}
components.  We choose to keep track of the quantities $\xi_+$
and $\xi_-$, which are defined below, as the two components of the
two-point function.  Therefore,
\beq
\bxi_2=(\xi_+, \xi_-)
\eeq

where

\beq
\xi_\pm=\frac{\txi_\pm}{\tw_2}
\eeq

and for all pairs of nodes, $A$ and $B$, with a separation distance
falling within the logarithmically binned configuration interval
$\eta$,

\beq
\txi_+(\eta)=\sum{\bar{\gamma_1}(A)\bar{\gamma_1}(B)+
      \bar{\gamma_2}(A)\bar{\gamma_2}(B)},
\eeq

\beq
\txi_-(\eta)=\sum{\bar{\gamma_1}(A)\bar{\gamma_1}(B)-
      \bar{\gamma_2}(A)\bar{\gamma_2}(B)},
\eeq

\beq
\tw_2(\eta)=\sum w(A)w(B).
\eeq

In these equations, ($\bar{\gamma_1}$, $\bar{\gamma_2}$) is the
weighted sum of ($\gamma_1'$, $\gamma_2'$)
for the galaxies contained in the node, weighted
by their inverse noise squared, and $w$ denotes the weight of the
node.
$\txi_\pm$ and $\tw_2$ are the raw correlation functions
which are binned by the pair separation.
$\xi_+$ and $\xi_-$ do not depend on the ordering of the two points but
only on the rotationally invariant spatial configuration which is
the separation distance between the two points in the case of the 2PCF.
The final two point function has two-components, and is stored
on a one-dimensional grid which corresponds to logarithmically binned
intervals of separation distances.


\section{Three-Point Correlation Function}
\label{sec:3PCF}

In weak lensing, the three-point function is an important tool
for detecting the non-Gaussianity in dark matter distribution,
which is in turn used to break the ${\Omega\--\sigma_8}$ degeneracy.
However, the three-point correlation function is computationally
more complex.
At face value, the 3PCF would appear to be an ${\cal O}(N^3)$
operation.  Such high computational cost makes it prohibitively
expensive to compute the full 3PCF for a large number of particles,
eg. $\sim 10^5$ galaxies that are in a single field of our current
lensing maps.
Fortunately, based on similar principles as the 2PCF computation,
we are able to compute the 3PCF in
${\cal O}(N\theta_c^{-4}\ln(\theta_c^2 N))$ time
(see section \ref{subsec:estimate_speed} for derivation).

\subsection{\bf Subdivision for 3PCF}
\label{sec:sub3}

Since we use a {\it binary} tree to compute the {\it three} point
function, the subdivision process can become tricky.
As in the subdivision process for the 2PCF described in
section \ref{sec:sub2},
we break up our discussion of the 3PCF subdivision into two parts:
the mainstream subdivision and the accuracy-dependent further
subdivision.
The algorithm flows only from the mainstream subdivision to the
further subdivision, but never in the opposite direction.
This is strictly a one-way traffic because the triplets of nodes,
which leave the mainstream subdivision, indicate areas amongst which
correlation statistics must be taken;
whereas, the further subdivision is used
only to correlate these nodes up to a certain accuracy level.

To summarize the whole subdivision process in a systematic way,
we construct a recursive subroutine named SUBDIVIDE3 with three
input arguments which are the indices of the three nodes
being compared, not necessarily distinct.
Let the current arguments be denoted by $k_1,
k_2$ and $k_3$, which are the indices of the nodes
$p_1, p_2$ and $p_3$ respectively;
and we write the subnodes of $p_i$ as $p_i^1$ and $p_i^2$,
with indices $k_i^1$ and $k_i^2$.
Similar to the 2PCF,
we only need to initiate the process by explicitly calling
SUBDIVIDE3($N+1,N+1,N+1$) once in the main program
where $N+1$ is the index of the root node;
this correlates among any three patches of area within the
root node.  Then the rest of the
processings are neatly taken care of by recursions.

\subsubsection{\bf Mainstream subdivision}
The mainstream subdivision ensures that
correlations are performed {\it on} and {\it amongst} all areas
of the map.
We again begin from the root node and move down the tree.
At the root level, when SUBDIVIDE3($N+1,N+1,N+1$) is invoked,
the three arguments correspond to a single patch of area,
thus we can only perform 'internal correlations'
by subdividing the 'root' node and comparing the subnodes.
However at the next level, two of the arguments may be distinct,
in which case we can do both 'internal' and 'external' correlations
by examining all possible combinations formed by any three
of the four immediate subnodes of the given pair of nodes.
Whenever we have three
distinct nodes to correlate amongst, these nodes are passed to
the further subdivision procedure.
We summarize all cases which may be encountered in the mainstream
subdivision.

{\it Case1} $\Leftrightarrow$ SUBDIVIDE3($k_1,k_2,k_3 | k_1=k_2=k_3$):
If we want to correlate within a node, as in the case of
the root node, we call SUBDIVIDE3 with three identical arguments.
This means that the input nodes are indeed the same.
In this case, there are four possible ways of choosing a triplet
of patches from its two subnodes, $p_1^1$ and $p_1^2$,
all of which must be explored:
\begin{description}
\item Choose all 3 points from $p_1^1$ if $k_1^1>N$\\
  $\Leftrightarrow$ SUBDIVIDE3($k_1^1,k_1^1,k_1^1$),
\item Choose all 3 points from $p_1^2$ if $k_1^2>N$\\
  $\Leftrightarrow$ SUBDIVIDE3($k_1^2,k_1^2,k_1^2$),
\item Choose 2 points from $p_1^1$ and 1 from $p_1^2$ if $k_1^1>N$\\
  $\Leftrightarrow$ SUBDIVIDE3($k_1^1,k_1^1,k_1^2$),
\item Choose 2 points from $p_1^2$ and 1 from $p_1^1$ if $k_1^2>N$\\
  $\Leftrightarrow$ SUBDIVIDE3($k_1^2,k_1^2,k_1^1$).
\end{description}
The first two possibilities correspond to correlations within
each of the subnodes while the last two indicate correlations
between the two subnodes.
Notice that in the last two possibilities, whenever SUBDIVIDE3 is
invoked with exactly two distinct arguments, the arguments are
ordered such that the first two are identical for coding simplicity.
Recursively calling SUBDIVIDE3 in this manner divides
the flow at each level.
Where the three arguments are identical,
as in the first two possibilities,
we repeat the same procedure as for a single node described above;
however, the last two possibilities lead us to {\it Case2}.

{\it Case2} $\Leftrightarrow$
SUBDIVIDE3($k_1,k_2,k_3 | k_1=k_2\neq k_3$):
When the arguments involve exactly two distinct nodes,
the first two arguments are always the same and the last different
by construction.
With this setup, we can easily determine the stage of flow by
checking the logical relations between adjacent arguments.
If $k_1=k_2$ and $k_2\neq k_3$,
this means that we intend to consider the correlation among any
three points with two points selected from $p_1$ and one from $p_3$.
Since the purpose of the mainstream subdivision is to specify
three distinct regions from where each point in a triplet
should be selected from, and
we must choose exactly one point from $p_3$, it makes no
difference other than increasing the computational cost to
further subdivide $p_3$, which corresponds
to specifying the region in $p_3$ from where that one point
is chosen from.
Whenever $p_1$ contains more than one particle, which is ensured
by construction, we subdivide it in each of the following three ways:
\begin{description}
\item Choose 2 points from $p_1^1$ and 1 from $p_3$ if $k_1^1>N$\\
   $\Leftrightarrow$ SUBDIVIDE3($k_1^1,k_1^1,k_3$),
\item Choose 2 points from $p_1^2$ and 1 from $p_3$ if $k_1^2>N$\\
   $\Leftrightarrow$ SUBDIVIDE3($k_1^2,k_1^2,k_3$),
\item Choose 1 point from $p_1^1$, 1 from $p_1^2$, and 1 from $p_3$\\
   $\Leftrightarrow$ SUBDIVIDE3($k_1^1,k_1^2,k_3$).
\end{description}
The subroutine SUBDIVIDE3 is recursively invoked with the appropriate
argument sets as indicated.
The first two possibilities bring us back to 
{\it Case2} while the third breaks into further subdivision
since three distinct nodes are specified.
Recursively performing the mainstream subdivisions ultimately covers
correlations over the entire map, and eventually
breaks down the 'root' node into distinct triplets of nodes which 
must be correlated amongst.  These triplets are subsequently delivered
to the further subdivisions.

\subsubsection{\bf Further subdivision}
{\it Case 3} $\Leftrightarrow$
SUBDIVIDE3($k_1,k_2,k_3 | k_1\neq k_2\neq k_3$):
If the arguments of the SUBDIVIDE3 subroutine are pairwise distinct,
the three input nodes are tested on if they are able to contribute
accurate enough three-point correlation statistics,
or need be further subdivided.
As in the 2PCF, the further subdivision does not call for the
collection of three-point statistics
until the 'open angle' criterion introduced in section
\ref{sec:theta} has been met by all three nodes.
The only subtleties are that the 'open angle' criterion must now
be satisfied by two ratios at each node, and that the possible
ways of failure are more varied for the 3PCF further subdivision.

Instead of checking all ratios for all three nodes every time the
further subdivision is called, we simply check the 'open angle'
criterion for the first node.
All other nodes are checked by calling further subdivision with
the input nodes in rotation once the current node has satisfied
the criterion.
We keep a counter $c$, initially set to zero, to
indicate the number of times the 'open angle' criterion has
been consecutively satisfied; when $c=2$, we call for
the node-to-node three-point correlation, shorthanded 3PC,
with the three distinct nodes as arguments.
Checking the criterion before each node is subdivided ensures
that only the minimal number of necessary further subdivisions
are performed.

Any three input nodes, $p_1$, $p_2$ and $p_3$, form a triangle,
non-degenerate or degenerate.
We shall call this triangle $\triangle ABC$, whose vertices are
labeled in the order of the input nodes,
and their opposite sides $a$, $b$ and $c$.
We let $r_1$ represent the size of $p_1$,
and let $p_1^1$ and $p_1^2$ denote its subnodes.
In further subdivisions, we must choose one point from each input node.

If $\max(\frac{r_1}{b},\frac{r_1}{c})\leq\theta_c$, then
$p_1$ satisfies the criterion and does not need be subdivided;
however, before we can call 3PC amongst these three nodes,
we must also check whether the other two nodes satisfy the 'open
angle' criterion if we have not already done so.
If any of the nodes in the triplet has not been tested 
$\Leftrightarrow c<2$, we call SUBDIVIDE3 with the nodes
in a rotated order such that in the
next round, we examine a different node which is labelled
$p_2$ currently.
Rotating the same triplets three times completes a full rotation.
Therefore, $c=2$ implies that the other two nodes in the current
configuration have already passed the 'open angle' test, and
since the current node also satisfies the criterion,
we can use this triplet to compute the 3PCF by calling the
subroutine 3PC where the raw correlation functions are accumulated.
This is put into pseudo-code below.\\
\\
In SUBDIVIDE3($k_1,k_2,k_3,c$) with $k_i=\delta_{ij} k_j$:\\
If $\max(\frac{r_1}{b},\frac{r_1}{c})\leq\theta_c$ or $k_1\le N$,
we have
\begin{verbatim}
  If c=2 then
    call 3PC(k1,k2,k3)
    return
  Else
    call SUBDIVIDE3(k2,k3,k1,c+1)
  Endif.
\end{verbatim}

If $\max(\frac{r_1}{b},\frac{r_1}{c})>\theta_c$ and $k_1>N$,
we subdivide $p_1$ by taking the following actions.
\begin{description}
\item Choose 1 point from $p_1^1$ and 1 each from $p_2$ and $p_3\\
   \Leftrightarrow$ SUBDIVIDE3($k_1^1,k_2,k_3,c=0$),
\item Choose 1 point from $p_1^2$ and 1 each from $p_2$ and $p_3\\
   \Leftrightarrow$ SUBDIVIDE3($k_1^2,k_2,k_3,c=0$).
\end{description}
We reset the counter $c$ to be zero in the case where a
different triplet configuration will be examined, for example,
when one of the nodes is subdivided.
When the triplet passes the 'open angle' test,
it is subsequently sent to the
node-to-node three-point correlation described in the next section.
In summary, the accuracy-dependent further
subdivision serves the sole purpose of
accumulating {\it accurate enough}
statistics amongst the three different area patches covered by the
triplets of distinct nodes leaving from the mainstream subdivision.

Similar to the 2PCF, this subdivision process will count all $\left(
\begin{array}{c}N\\3\end{array}\right)$
triplets when $\theta_c$ is set to zero.
This is a ${\cal O}(N^3)$ process as in the computation of the 3PCF
by definition, and is prohibitively
expensive for $N$ of astronomical interest.
The accuracy parameter $\theta_c$ controls the depth of subdivision and
makes this otherwise intractable calculation feasible.
One expects the truncation error from tree walking
to scale as $\theta_c^2$.
For a comparison between the 3PCF computed using our fast algorithm
with nonzero $\theta_c$
and the fully accurate results obtained using the brute force approach,
refer to section \ref{sec:speed}.

\subsection{\bf Node-to-node three-point correlation in 2D}
\label{sec:nn3}

The subroutine 3PC computes the raw correlation functions for
the weighted field quantities and for the weights using
triplets of distinct nodes that exit the subdivision procedure.
These correlation functions are then used to compute the
final 3PCF which is the quotient of the two raw functions.
Here, we describe the node-to-node three-point correlation
procedure for both 2D scalar fields and spin-2 fields.

We consider the information about each node,
obtained while building the tree,
to be concentrated at its weighted centre of mass; thus, we may
consider the nodes point-like in the following discussion.

\subsubsection{\bf 3PC for a 2D scalar field}
\label{sec:3scalar}

The 3PCF for a scalar field is a scalar function of
triangle configurations, defined on a three-dimensional grid.
The 3PC for a scalar field relies on a map from the space of triangles
to the three-dimensional configuration space,
which does not assume parity,
as a parameterization of triangle configurations.
Here, we describe the procedure of node-to-node three-point
correlation for the scalar field $\kappa$.

We measure the position of the galaxies in Cartesian coordinates,
their projected mass density $\kappa$ and the noise.
For any triplet $\{p_1,p_2,p_3\}$ entering the 3PC subroutine,
we define the sides $a$, $b$, $c$, and their opposite
vertices $A$, $B$, $C$ for the triangle
formed by the triplet as follows.
Given any three points, we first identify the longest side and name
it $a$.  We then name the other two sides $b$ and $c$ in the
counter-clockwise direction.

For numerical reasons, we discretize the finite region of the
configuration space accessible to triplets in the system
into finitely many $3$-dimensional rectangles $\eta$
by logarithmically binning each of the length dimensions.
Therefore, the three-point function $\xi_3^{\kappa}$ is given by
the quotient of the raw correlation functions as
\beq
\xi_3^{\kappa}=\frac{\txi_3^{\kappa}}{\tw_3}
\eeq
where, for all triplets $\{A, B, C\}$ with coordinates
($s_1$, $s_2$, $s_3$) $\in\eta$ in the configuration space,
which are the lengths of the sides $a$, $b$ and $c$ repectively,
and $\bar{\kappa}$ the weighted sum of $\kappa$ at the node,
\beq
\txi_3^{\kappa}(\eta)=\sum{\bar\kappa(A)\bar\kappa(B)\bar\kappa(C)}
\eeq
and
\beq
\tw_3(\eta)=\sum{w(A)w(B)w(C)}.
\eeq

In weak lensing, $\kappa(\bX)$ is the projected matter overdensity.
To calculate the skewness in the projected dark matter distribution
from the three point function of $\kappa$, $\xi_3^{\kappa}$,
we adopt the compensated Gaussian filter ${\cal U}$ used in
\citet{2003ApJ...592..664P} as the smoothing window function for
$\kappa$, which gives
\beq
\langle\bar{\kappa}^3\rangle=2\pi\int\xi_3^{\kappa}(r,\bX)U_3(r,\bX)rdrd^2x.
  \label{eqn:kappa3}
\eeq
To accommodate our choice of coordinates, with
$(s_1,s_2,s_3)$ being the length of $a$, $b$ and $c$ respectively,
we obtain
\begin{eqnarray}
\langle\bar{\kappa}^3\rangle&=&
4\pi\int_0^{\infty}\int_0^{\infty}\int_{|s_1-s_2|}^{|s_1+s_2|}
\xi_3^{\kappa}(s_1,s_2,s_3)U_3(s_1,s_2,s_3)\nonumber\\
&\times&\frac{s_1 s_2 s_3 ds_3 ds_2 ds_1}
{\sqrt{-s_3^4-s_2^4-s_1^4+2 s_2^2 s_3^2+2 s_1^2 s_3^2+2 s_1^2
s_2^2}}.
\end{eqnarray}

\subsubsection{\bf 3PC for a 2D spin-2 field}

The 3PCF for a spin-2 field is conventionally a function
with eight components defined on a three-dimensional
configuration space.
In addition to the parameterization of triangles,
there is an orthogonal transformation involved in the 3PC
for a spin-2 field.

We measure the position of the galaxies in Cartesian coordinates,
their projected mass density $\kappa$ and the noise.
The three-point correlation statistics for each triplet is
obtained after a coordinate change for the {\it ordered} triplet,
hence it is dependent upon the ordering of the points.
To reduce the complexity of the problem,
for any triplet $\{p_1,p_2,p_3\}$ entering the 3PC subroutine,
we define the sides $a$, $b$, $c$, and their opposite
vertices $A$, $B$, $C$ for the triangle
formed by the triplet in the same way as
in section \ref{sec:3scalar} for a scalar field.
Recall that we define $\triangle ABC$ by first identifying the longest side
as $a$, and the other two sides $b$ and $c$ in the counter-clockwise direction.

Similarly, we choose ($s_1$, $s_2$, $s_3$), which are the lengths of
$a$, $b$ and $c$ repectively,
as the coordinates in the configuration space
while each spatical dimension is binned logarithmically
to give finitely many $3$-dimensional rectangles $\eta$
in the configuration space.

Since we are interested in the 3PCF for the spin-2 field
relative to the configuration rather than to any fixed spatial
coordinates,
we need to project $\gamma$ onto a coordinate system
intrinsic to the ordered triplet $\{A,B,C\}$.
Let $\xp$ be the unit vector ${\bmath\widehat{BC}}$, and
$\yp$ the unit vector which is perpendicular to $\xp$ and points
from line segment connecting $B$ and $C$ to the point $A$.
We want to obtain an expression for $\gamma$ in terms of these
new coordinates.
Suppose $\gamma$ is a spin-1 object, then the projections of $\gamma$
onto $\xp$ and $\yp$ are simply $\gamma_x'$ and $\gamma_y'$
respectively with
\beq
\gamma_x'=\gamma\cdot\xp
\eeq
and
\beq
\gamma_y'=\gamma\cdot\yp.
\eeq

However, our $\gamma$ is a spin-2 object.  To compute its projection,
it is necessary to introduce the matrix $\bGamma$ defined as
\beq
\bGamma=\left(\begin{array}{cc}
\gamma_1 & \gamma_2\\
\gamma_2 & -\gamma_1
\end{array}\right).
\eeq
The projected shear is given by $\gamma'=(\gamma_1',\gamma_2')$ where
\beq
\gamma_1'=\frac{1}{2}((\xp)^{T}\bGamma\xp - (\yp)^{T}\bGamma\yp)
\eeq
and
\beq
\gamma_2'=\frac{1}{2}((\xp)^{T}\bGamma\yp + (\yp)^{T}\bGamma\xp).
\eeq
Here, $\xp$ and $\yp$ are taken to be unit {\it column} vectors.
Notice that $|\gamma'|=|\gamma|$
since the transformation is orthogonal.

Recall that the weight of each node, $w$, is the sum of
of the weights of all galaxies in the node; and the weight of a galaxy
is its inverse noise squared.
We compute the raw three point functions for the weight, $\tw_3$,
as well as for the eight components of the 3PCF,
$\txi_{111}$, $\txi_{112}$, $\txi_{121}$, $\txi_{211}$,
$\txi_{122}$, $\txi_{221}$, $\txi_{212}$ and $\txi_{222}$, defined as
\beq
\tw_3(\eta)=\sum w(A)w(B)w(C)
\eeq
and
\beq
\txi_{ijk}(\eta)=\sum\bar{\gamma_i}(A)\bar{\gamma_j}(B)\bar{\gamma_k}(C)
\eeq
for all triplets $\{A,B,C\}$ belonging to the set of
configurations $\eta$, that is $(s_1,s_2,s_3) \in \eta$.
In the equations, ($\bar{\gamma_1}$, $\bar{\gamma_2}$) represent the
weighted sum of ($\gamma_1'$, $\gamma_2'$) at the node.
The final three-point correlation function $\bxi_3$
with eight components is the quotient
of the raw correlation functions $\txi_{ijk}$ for the components
and $\tw_3$ for the weight, that is

\beq
\bxi_3=\{\xi_{ijk}\}
\eeq

with each component

\beq
\xi_{ijk}=\frac{\txi_{ijk}}{\tw}.
\eeq

The 3PCF for a two-dimensional spin-2 field is a eight-component
function defined on a three-dimensional grid corresponding
to the logarithmically binned configuration space $(s_1,s_2,s_3)$.


\section{$\bmath{n}$-point correlation function in
         $\bmath{k}$ dimensions}
\label{sec:nPCF}

The algorithm for the 2PCF and the 3PCF can be easily extended
to $n$-th order.
In this section,
we discuss how the $n$-point correlation function can be computed
for a $k$-dimensional scalar field.
The binary tree construction is entirely independent of the
order of the correlation function.
Although the subdivision process needs be generalized for
higher order correlation functions (ie. n-dependent), its structure
is independent of the field dimension (ie. k-independent).
The k-independence of the subdivision is supported by the
fact that once the tree has been constructed, the subdivision process
relies on the {\it binary} structure of the tree rather
than on the spatial struture of the particles.
The node-to-node $n$-point correlation ($n$-PC) for a scalar field
relies on a unique counting of $n$-sided polygons in a
$k$-dimensional space.

\subsection{\bf Build tree in $\bmath{k}$ dimensions}
In a k-dimensional space, the tree construction is virtually the
same as that in 2D.
At each node $p$ which contains more than one particle,
we find the weighted centre of mass, $\bX_c$,
and determine among all particles in the node the one that is
furthest from $\bX_c$, ie. the particle with position
vector $\bX_{max}$ such that
$\|\bX_{max}-\bX_c\|=Sup\|\bX-\bX_c\| \mbox{ for all } \bX \in p$.
We then bisect the node by cutting along a $(k-1)$-dimensional
subspace, or affine space, which is perpendicular to the line
connecting $\bX_{max}$ and $\bX_c$,
and through a projected median point on this line determined
as in section \ref{sec:buildtree}.
Similar to the 2D case, the binary tree remains useful
for computing the $n$-point correlation function
where $n$ is any positive integer no greater than $N$.

\subsection{\bf Subdivision for $\bmath{n}$-PCF}
The subdivision process for computing the $n$-point correlation
function can be divided into two parts as with the 2PCF and the 3PCF,
and be dealt with in a single recursive subroutine SUBDIVIDE
which takes n arguments.
The mainstream subdivision occurs when
the input nodes are not pairwise distinct; otherwise, the
$\theta_c$-dependent further subdivision is performed.
The subroutine SUBDIVIDE({\bnodes}, $n$, $c$),
where {\bnodes} is a 1D array
of length $n$ containing the indices of the $n$ input nodes,
is discussed below.

\subsubsection{\bf Mainstream subdivision}
$\Leftrightarrow$
SUBDIVIDE($k_1,...,k_n|k_i=k_j\mbox{ for some }i\neq j$):\\
If the $n$ elements of {\bnodes} are not pairwise distinct, we
want to call SUBDIVIDE with all possible combinations, formed by
the non-repeating nodes and the subnodes of the repeating ones, as
input argument sets.
Interested readers should see the appendix for an
outline of the n-PCF mainstream subdivision scheme.
We call SUBDIVIDE in such a way as to keep the repeating nodes
together.
This specific ordering allows us to easily determine whether a set
of nodes is pairwise distinct and find the repeated ones by
checking only the adjacent nodes.

\subsubsection{\bf Further subdivision}
$\Leftrightarrow$
SUBDIVIDE($k_1,...,k_n|k_i\neq k_j\mbox{ for any }i\neq j$):\\
When all $n$ input nodes are pairwise distinct, further
subdivision of the first node $p_1$ is performed upon
the condition that the 'open angle'
criterion is not satisfied by $p_1$.
The 'open angle' criterion for $p_1$, in the n-th order case, is the
condition that $\max(\frac{r_1}{d_i})\leq\theta_c$ where $r_1$ is
the size of $p_1$, and $d_i$ for $i\in\{2,...,n\}$, is the
separation distance between $p_1$ and $p_i$.
Whenever $p_1$ meets the criterion, we check whether
the counter $c$, which keeps track of the number of points
on the $n$-sided polygon that have passed the 'open angle' test,
reads $n-1$.  If so, we accumulate $n$-point statistics using 
these nodes by calling the subroutine $n$-PC; otherwise,
we recursively call SUBDIVIDE with the input nodes rotated by one.
This process is demonstrated below.\\
\\
In SUBDIVIDE(${\bmath nodes},n,c$) with $k_i=\delta_{ij} k_j$:\\
If $\max(\frac{r_1}{d_i})\leq\theta_c$ or ${\bmath nodes}(1)\le N$,
we have
\begin{verbatim}
  If (c=n-1) then
    call n-PC(nodes,n)
    return
  Else
    call SUBDIVIDE(cshift(nodes,1),n,c+1)
  Endif.
\end{verbatim}

If $p_1$ does not satisfy the 'open angle' criterion and it
contains distinct subnodes,
we {\it replace} the correlation amongst the $n$ input nodes by
a pair of correlations each accounting for one of the combinations
formed by a subnode of $p_1$ and the remaining $n-1$ nodes.
This is summarized in code format below.\\

In SUBDIVIDE(${\bmath nodes},n,c$) with $k_i=\delta_{ij} k_j$:\\
If $\max(\frac{r_1}{d_i})>\theta_c$ and ${\bmath nodes}(1) > N$,
\begin{verbatim}
  snodes=nodes
  snodes(1)=nodes(1).sub1
  call SUBDIVIDE(snodes,n,c=0)
  snodes(1)=nodes(1).sub2
  call SUBDIVIDE(snodes,n,c=0).
\end{verbatim}

We reset the counter $c$ to zero in the case where one of the nodes
must be subdivided since a new $n$-tuple configuration
is subsequently considered.
This way, the counter reads $n-1$ if and only if all the other
$n-1$ nodes besides the one being examined
have satisfied the 'open angle' criterion.

\subsection{\bf Node-to-node $\bmath{n}$-point correlation for scalar fields}
This section includes
a general discussion about the parameterization
of $n$-tuple configurations in $k$ dimensions (section \ref{sec:nkd})
and
a detailed description of how the $n$-PC
can be done for a two-dimensional scalar field (section \ref{sec:n2d}).

\subsubsection{\bf Node-to-node $\bmath{n}$-point correlation of a
$\bmath{k}$-dimensional scalar field}
\label{sec:nkd}

Let us consider a single point in a $k$-dimensional space, the point
itself is dimensionless and rotationally invariant;
therefore, when a second point is placed in the system,
only its distance from the first point matters.  When we add a
third point, for $k\geq 2$, two parameters are generally required
to specify its position relative to the first two points which
form a one dimensional subspace;
one of the parameters can be its orthogonally projected position onto
the line segment connecting the first two points,
and the other can be its
perpendicular distance from the line.
Given three points, they may form a plane; for $k\geq 3$,
specifying yet another point relative to them requires
three parameters; a set of parameters could be the projected
position of the fourth point onto the plane (2 parameters), and its
distance from the plane (1 parameter).
However, a point in a $k$-dimensional space cannot have
more than $k$ independent components, which sets an upper bound
for the number of parameters.
In general, before the $i^{th}$ point is placed in a
$k$-dimensional space, there pre-exists $i-1$ points,
potentially forming an object of dimension $\min(i-2,k)$;
the position of the $i^{th}$ point in relation to the pre-existing $i-1$
points can be specified by its orthogonal projection onto and its
distance from the object formed by the first $i-1$ points;
this requires no more than $\min(i-1,k)$ parameters.  
Consider forming an $n$-sided polygon in $k$-dimensional space
by adding one point after another, we see that the $n$-sided
polygon can be uniquely specified, up to rotation, by no more than
$m$ parameters where
\beq
m=\sum_{i=1}^n \min(i-1,k).
\eeq
Hence, the full $n$-point correlation function in $k$-dimension
can be stored on a grid of dimension $m$ defined as above.

\subsubsection{\bf Node-to-node $\bmath{n}$-point correlation of
a 2D scalar field}
\label{sec:n2d}

To compute the $n$-point correlation function for a 2D scalar field,
we follow the tree construction and subdivision
as described previously.  Here, we discuss how one handles the
statistics for a given set of $n$ distinct points \{$p_1,...,p_n$\}.
We first dismiss the $n$-tuples with anomaly such as the coincidence
of two or more points; thus, we need not consider this case
in the following discussion.


We wish to find a parameterization that uniquely maps any $n$-tuple
\{$p_1,...,p_n$\} to a point in the $m$-dimensional configuration space,
such as the parameterization suggested in section \ref{sec:nkd},
with
\beq
m=\sum_{i=1}^n \min(i-1,2)
\eeq
which, for $n > 1$, reduces to
\beq
m=2n-3.
\eeq

For $n$ points in two dimensions, we may order the points and parameterize
the configuration as follows.  We first define line $l$ as the line
joining the two points of widest separation.  Let $\yp$ be the unit
vector perpendicular to line $l$ and pointing towards the side with the
greater number of points.  We define $\xp$ as the unit vector orthogonal
to $\yp$ such that $\xp$ and $\yp$ form a right-handed system.
We call the point on line $l$ with the smaller x'-component $p'_1$
and translate the coordinate system so that $p'_1$ coincides with
the origin.  The point on line $l$ with the larger x'-component and
coordinates ($x'_2$, $0$) in the new coordinate system is then named $p'_2$.
We order the other $n-2$ points by increasing angle from the
${\bf x'}$-axis.  We now have an ordered $n$-tuple \{$p'_1,...,p'_n$\} with
new coordinates \{($x'_i$, $y'_i$), i=1,...,n\}, amongst which $x'_1$,
$y'_1$ and $y'_2$ are zero.
The ordered $2n-3$ numbers ($x'_2$, $x'_3$, $y'_3$, ..., $x'_n$, $y'_n$)
parameterize the $m$-dimensional configuration space, and are unique up to
translations and rotations of the configuration.

Due to discrete nature of data storage, we partition the finite
region of the $m$-dimensional continuous configuration space accessible
to the $n$-tuples in the system, into finitely many $m$-dimensional
rectangles $\eta$'s.

For all ordered $n$-tuples \{$p_1',...,p_n'$\} $\in \eta$,
we can compute the raw $n$-point function $\txi_n$
for the scalar field $\rho$, and $\tw_n$ for the weight,
as a function of discretized configuration, by
\beq
\txi_n(\eta)=\sum(\ \prod_{i=1}^n \bar{\rho}(p_i')\ ),
\eeq
and
\beq
\tw_n(\eta)=\sum(\ \prod_{i=1}^n w(p_i')\ )
\eeq
where $\bar{\rho}$ is the weighted sum of $\rho$ among the
particles in the node.

The final $n$-point correlation function $\xi_n$ is a scalar
function on an $m$-dimensional grid given by
\beq
\xi_n(\eta) \equiv \langle \prod_{i=1}^n \rho(p_i^{\eta}) \rangle
=\frac{\txi_n(\eta)}{\tw_n(\eta)}.
\eeq
Here, $\{{\bmath p}^{\eta}\} =
\{(p_1^{\eta},p_2^{\eta},...,p_n^{\eta})\}$ is the set of
all $n$-tuples which are mapped to the $m$-dimensional rectangle
$\eta$ in the configuration space under the specified parameterization.


\section{Accuracy and Speed in 2D}
\label{sec:speed}

The accuracy and speed of our algorithm in 2D
are analyzed in this section.
We first show that tests of algorithm accuracy for the 3PCF
yield results which are consistent with our expectation.
We then turn to examine the algorithm speed and memory usage
of the 2PCF and 3PCF computations,
with focus on the computational costs of the subdivision
and node-to-node correlation procedures.
To do this, we offer a theoretical estimate of the computational
costs, followed by results and analysis from various tests
of algorithm speed.  Finally, we compare the costs of our 2PCF
and 3PCF algorithms with that of other correlation algorithms.

\subsection{\bf Tests of Accuracy for 3PCF}
\label{subsec:test_accuracy}

The error of the algorithm, due to the truncation error of
the binary tree, is expected to scale as $\theta_c^2$.
Here, $\theta_c$ is an accuracy parameter specifying the
critical linear angle, as seen from a node
$p_i$, above which size another node $p_j$ cannot
be treated as a unit and must be subdivided.

We test the accuracy of the algorithm in 2D 3PCF computation for
$N=1000$ using CITA's 1.3GHz ItaniumII, Dell Poweredge 7250 computer.
For this set of computations, we make use of a mock catalogue of
galaxies with $x$, $y$ coordinates, $\gamma_1$, $\gamma_2$,
$\kappa$ and noise.
The positions of the galaxies are randomly generated with
coordinate components taking values between $0$ and $5$.
Since $\gamma_1$ and $\gamma_2$ are not used in accuracy testing,
we set them to zero.  For equal weighting in the 3PCF, the noise
figures of the galaxies are identically assigned the value of $1$.
The scalar field $\kappa$ assumes a two-dimensional Gaussian
distribution around the centre of the map given by
\begin{equation}
  \kappa(x,y)=\exp\left[-\frac{(x-x_0)^2+(y-y_0)^2}{2 \sigma^2}\right].
\end{equation}
Here, ($x_0$, $y_0$)=($2.5$, $2.5$) are the coordinate components of the 
map centre, and the width of the Gaussian distribution, $\sigma=1.25$,
is a quarter of the map's side length.
The 3PCF is binned at logarithmic intervals of $2^{0.5}$
over the length scale from $0.1$ to $\sqrt{2}\times 5$.
The higher cutoff length corresponds to the diagonal length of the
2D lensing map.

\begin{figure}
\vskip 3.3 truein
\includegraphics{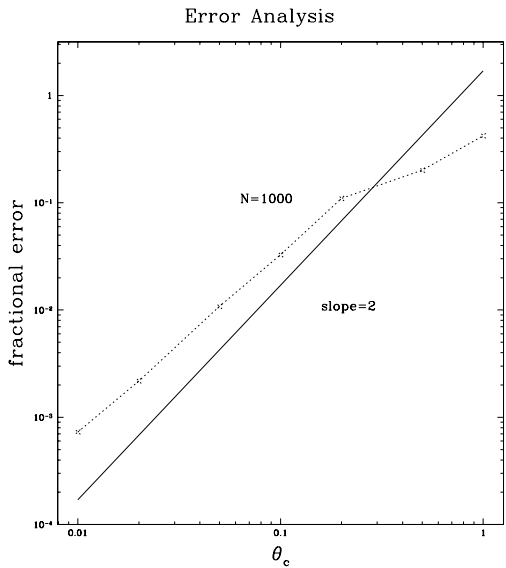}
\caption{For $1000$ particles, we plot the fractional error
of the 3PCF for a scalar field computed using the new algorithm
compared to that using the brute force approach.
Recall that the 3PCF is stored on a 3D grid.
We first smooth over the neighbouring grids, then compute the
fractional error
$\Delta=\sqrt{\frac{\sum_{i,j,k}(\xi_{\theta_c}-\xi_{\ast})^2}{
\sum_{i,j,k}\xi{\ast}^2}}$ where $\xi_{\ast}$ is the 3PCF computed
by direct summation over all triplets of particles.
We see that the truncation
error scales as $\theta_c^2$ which agrees with our estimate.}
\label{fig:accuracy}
\end{figure}

The truncation error for the 3PCF computation on a 2D scalar field
using the tree approach is plotted against
different $\theta_c$ in figure \ref{fig:accuracy}.
Empirical runs show that the error is
proportional to ${\theta_c}^{2}$ as predicted in section
\ref{sec:sub3}.
When $\theta_c=0$, the fractional error is around the order of
$10^{-15}$, and is therefore neglectable.

In tree-based $N$-body simulations, typical values of the critical
open angle are $0.5<\theta_c<1$.  \citet{1989ApJS...70..389B}
find that the violation of energy conservation in a simulation
using a tree-based force computation is only $2\%$ for $\theta_c=1$.
Although the fractional error in $\langle\bar{\kappa}^3\rangle$
differs from the fractional violation of energy conservation
by definition,
and that the procedures used to obtain them are quite different,
the high accuracy for large $\theta_c$ in $N$-body simulations
suggests a similar relationship in our case.
The quantity of interest, $\langle\bar{\kappa}^3\rangle$,
may turn out to be accurate enough for our purpose even
when $\theta_c$ is large.
Furthermore, only a small subset of all triangle configurations
are ever considered for most scientific questions derived from
the 3PCF.
As a result, the errors in the smoothed skewness fields, eg. equation
\ref{eqn:kappa3}, derived from the partial 3PCF are probably
comparable to that in figure \ref{fig:accuracy}.
Hence, it is possible that values such as $\theta_c=0.5$ are quite
sufficient for 3PCF computations.

\subsection{\bf Estimate of Computational Cost for 2PCF and 3PCF}
\label{subsec:estimate_speed}

Here, our primary goal is to provide a theoretical estimate for the
cost of the 3PCF algorithm in 2D.
In addition, we derive the computational cost of the 2D 2PCF
at the end of the subsection.

In the brute force 3PCF computation, the computing time can be broken
down into the following components, reading of the catalogue,
computation of the raw correlation functions, normalization of
the raw correlation functions for $\gamma$ and $\kappa$ by the
raw weight correlation function, and finally the output of the
full 3PCF.
In addition to the above procedures for brute force calculations,
the 3PCF computation using
our fast algorithm also spends time on tree building
just after reading the catalogue.
Among these processes, the reading of catalogue and the output of
3PCF are considered extrinsic to the computation algorithm.
In a gridded 3D configuration space, the normalization cost
is independent of $\theta_c$ and $N$.
Here, we only discuss the speed of tree
building and that of the raw correlation function computation.

In 2D, the construction of the tree costs ${\cal O}(N (\log N)^2)$
for $N$ particles.
In the $k$-dimensional case, however, building the tree costs
${\cal O}(k N (\log N)^2)$ because it involves solving
$k$ components of the distances.
Since the cost of the tree construction is relatively well-understood,
and that its prefactor is small (see figure \ref{fig:speed_n}),
we will focus on the cost of raw 3PCF computation in the
remaining section.

In the following paragraphs, we offer a derivation for the
computational cost of the raw 3PCF.
We show that when $\theta_c^2 N$ is sufficiently large, the computing
time scales as $N\theta_c^{-4} \ln(\theta_c^2 N)$.
This slowly converges to ${\cal O}(N \theta_c^{-4})$
as $\theta_c^2 N$ gets very large.

Since the isosceles and equilateral triangles constitute a set of
measure zero in the configuration space, we can neglect them from
the integral used to estimate the total number of contributing
triplets\footnote{We use the term
'contributing triplets' in place of 'triplets which contribute to the
correlation statistics.}.
Here, we consider an arbitrary triplet of nodes forming a scalene
triangle.
We call the node opposite to the longest side $p_1$, the node closer
to $p_1$ $p_2$, and the other node $p_3$.  We then label the side
connecting $p_i$ and $p_j$ as $d_{ij}$, and we have $d_{ij}=d_{ji}$.
Furthermore, we denote the size of $p_i$ by $r_i$.

In the given triplet configuration, since $d_{12}<d_{13}<d_{23}$,
the open angle criterion demands
\beq
  \frac{r_1}{d_{13}}<\frac{r_1}{d_{12}}<\theta_c,
  \label{eqn:r1}
\eeq
\beq
  \frac{r_2}{d_{23}}<\frac{r_2}{d_{12}}<\theta_c
\eeq
and
\beq
  \frac{r_3}{d_{23}}<\frac{r_3}{d_{13}}<\theta_c.
\eeq
We are only interested in triplets which would contribute to the
correlation statistics.
For such a triplet, if any of the nodes were replaced by its
parent node,
the obtained triplet would no longer satisfy the open angle criterion.
Except at the bottom of the tree, the area of the parent node is
expected to double that of the current node, hence the expected
increment in linear node size one level up gives a factor of $\sqrt{2}$.
Therefore, we obtain for contributing triplets
\beq
  \frac{r_1}{\theta_c}<d_{12}<\sqrt{2}\frac{r_1}{\theta_c},
  \label{eqn:d12_r1}
\eeq
\beq
  \frac{r_2}{\theta_c}<d_{12}<\sqrt{2}\frac{r_2}{\theta_c}
\eeq
and
\beq
  \frac{r_3}{\theta_c}<d_{13}<\sqrt{2}\frac{r_3}{\theta_c},
\eeq
which then yields
\beq
  \frac{1}{\sqrt{2}}\theta_c d_{12} < r_1 < \theta_c d_{12},
\eeq
\beq
  \frac{1}{\sqrt{2}}\theta_c d_{12} < r_2 < \theta_c d_{12}
\eeq
and
\beq
  \frac{1}{\sqrt{2}}\theta_c d_{13} < r_3 < \theta_c d_{13}.
\eeq
We see that the sizes of $p_1$ and $p_2$ are governed by $d_{12}$
whereas the size of $p_3$ is governed by $d_{13}$.

Let us now fix an arbitrary node in the tree hierarchy as $p_1$,
and find another pair of nodes $p_2$ and $p_3$ such that
they form a contributing triplet with $d_{12}<d_{13}<d_{23}$.
We set up consecutive annuli centred at $p_1$, outlined by
concentric circles of radius which increases with a factor of
$\sqrt{2}$, starting with the annulus given by
$\frac{r_1}{\theta_c}<d<\sqrt{2}\frac{r_1}{\theta_c}$
where $d$ is the distance from $p_1$.

We argue in the following that the possibility of finding a node
in any of these annuli, which qualifies as a member of a contributing
triplet containing $p_1$ in the manner described above,
is proportional to $\theta_c^{-2}$.
The area of the $j^{th}$ annulus with $d_j<d<d_{j+1}$ where
$d_j=2^{\frac{j-1}{2}}\frac{r_1}{\theta_c}$ is given by
\beq
\pi (d_{j+1}^2 - d_j^2) = \pi d_j^2;
\eeq
whereas, a qualifying node\footnote{We use the term 'qualifying
node' in place of 'a node which qualifies as a member of a
contributing triplet containing $p_1$ such that $p_1$ is the
node opposite to the longest side of the triangle'.}
in the $j^{th}$ annulus has an area
proportional to $\theta_c^2 d_j^2$.
Furthermore, nodes of similar sizes should not overlap.
Thus, the possibility of finding a qualifying node in any of the
annuli is $\theta_c^{-2}$ up to a constant factor which is
independent of $j$.

As shown above, in a contributing triplet,
the distance of $p_2$ from $p_1$ in terms of $r_1$ is given
by equation \ref{eqn:d12_r1}.
From equation \ref{eqn:r1} and the definition of $L$ as the
largest node separation in the image,
the distance of $p_3$ from $p_1$ can also be expressed in terms
of $r_1$ as
\beq
\frac{r_1}{\theta_c}<d_{13}<L.
\eeq
Geometrically speaking,
for a node to qualify as $p_2$ in such a configuration, it must fall
in the first annulus.
The number of different nodes, which qualify to be $p_2$, is thus
proportional to the possibility of finding a qualifying node in the
first annulus, that is $\propto \theta_c^{-2}$ as derived above.
However, $p_3$ can fall in any of the annuli, with each annulus
containing roughly $\theta_c^{-2}$ qualifying nodes.
Since there are about $\ln \frac{L}{r_1/\theta_c}$ such annuli in
a map with a maximum node separation $L$,
the number of different nodes which qualify to be $p_3$
is proportional to $\theta_c^{-2} \ln \frac{L \theta_c}{r_1}$.
The total number of contributing triplets,
$M_3$, can then be estimated by
\beq
  M_3\propto \int_{r_{min}}^{r_{max}} \frac{dN_1}{dr_1}
             \frac{1}{\theta_c^4} \ln\frac{L \theta_c}{r_1} dr_1
\eeq
where $\frac{dN_1}{dr_1}$ is the number of nodes with size
between $r_1$ and $r_1+\Delta r_1$ divided by $\Delta r_1$
in the limit that $\Delta r_1 \rightarrow 0$.
Since $dN_1\propto \frac{L^2}{\pi r_1^2}$ and $dr_1\propto r_1$,
we have
\beq
  \frac{dN_1}{dr_1} \propto \frac{L^2}{\pi r_1^3}.
\eeq
Therefore,
\begin{eqnarray}
  M_3&\propto& \frac{L^2}{\pi \theta_c^4} \int_{r_{min}}^{r_{max}}
               \frac{1}{r_1^3} \ln\frac{L \theta_c}{r_1} dr_1 \\
     &=& \frac{L^2}{2\pi \theta_c^4} \left[ \frac{1}{2 r_1^2}
         - \frac{1}{r_1^2} \ln\frac{L\theta_c}{r_1}
         \right]_{r_{min}}^{r_{max}}.
\end{eqnarray}
Substituting in $r_{min}\sim \sqrt{\frac{L^2}{\pi N}}$
and $r_{max}\sim L\theta_c$, we obtain
\begin{eqnarray}
  M_3&\propto& \frac{L^2}{2\pi\theta_c^4}\left[\left(\frac{1}{2 r_{max}^2}
               - \frac{1}{r_{max}^2} \ln\frac{L\theta_c}{r_{max}}\right)
               - \left(\frac{1}{2 r_{min}^2} - \frac{1}{r_{min}^2}
               \ln\frac{L\theta_c}{r_{min}} \right) \right] \nonumber\\
    &\sim& \frac{L^2}{2\pi\theta_c^4}\left[\left(\frac{1}{2 L^2\theta_c^2}
           \right) - \left(\frac{\pi N}{2 L^2} - \frac{\pi N}{L^2}
           \ln\frac{L\theta_c}{\sqrt{L^2/\pi N}} \right) \right] \\
    &=& \frac{N}{4 \theta_c^4} \left[ \left(\frac{1}{\pi\theta_c^2 N}
        \right) - 1 + 2
        \ln\left(\theta_c\sqrt{\pi N}\right) \right] \\
    &=& \frac{N}{4 \theta_c^4} \left[ \left(\frac{1}{\pi\theta_c^2 N}
        \right) - 1 + \ln\pi +
        \ln\left(\theta_c^2 N\right) \right] \\
    &\simeq& \frac{N}{4 \theta_c^4} \left[ \left(\frac{1}{\pi\theta_c^2 N}
        \right) +0.14 +
        \ln\left(\theta_c^2 N\right) \right].
\end{eqnarray}
In the limit $\theta_c^2 N \gg 1$, we have
$\frac{1}{\pi\theta_c^2 N} < \frac{1}{\theta_c^2 N} \ll 1$ and
$\ln\left(\theta_c^2 N\right) \gg 1$.  Hence,
\beq
  M_3\propto \frac{N}{\theta_c^4} \ln\left(\theta_c^2 N\right).
\eeq
The cost of the computing the raw 3PCF is proportional to the
number of contributing triplets.
Hence, the computational cost of the raw 3PCF scales as
$\frac{N}{\theta_c^4} \ln\left(\theta_c^2 N\right)$
for sufficiently large $\theta_c^2 N$.
As $\theta_c^2 N \rightarrow \infty$, $\ln\left(\theta_c^2 N\right)$
can be considered as a constant since it increases only slowly.
Hence, in this limit,
\beq
  M_3\propto \frac{N}{\theta_c^4}.
\eeq
However, this convergence of the computational cost to
${\cal O}(N \theta_c^{-4})$
behaviour as $\theta_c^2 N$ increases is very slow.


In the limiting case where $\theta_c\rightarrow 0$,
the relation that the expected number of nodes in each annulus is
proportional to $\theta_c^{-2}$ no longer holds.
This is because $\theta_c^{-2}\rightarrow\infty$
while the number of non-overlapping nodes in the area remains
bounded above by the total number of particles $N$.
From the tree's perspective, subvisions descend all the way
down to the leaves for small $\theta_c$,
and perform full correlations among all triplets of particles,
conforming to the intention of the subdivision process.
On the other hand, as $N\rightarrow 0$ with a fixed $\theta_c$,
the subdivision procedure easily reaches the bottom of the tree
due to the shallowness of the tree.  Hence, the node-to-node
correlations tend to be done at the leaf level
for both small $\theta_c$ and small $N$.
At the leaf level where $p_1$ contains a single galaxy,
any pair of galaxies qualifies to form a contributing triplet
with $p_1$ since, for the open angle criterion,
the size of a single galaxy is extremely small
compared to the space between galaxies.
The possibility of finding a galaxy which qualifies to be a member
of a contributing triplet containing $p_1$ is thus $N-1$.
In the limit where $\theta_c^2 N \ll 1$,
the total number of triplets is $N(N-1)(N-2) \sim N^3$,
so is the scaling of the computational cost for subdivisions
and node-to-node correlations.
Taking the limit as $\theta_c$ goes to $0$, the
$\theta_c$-independent scaling of ${\cal O}(N^3)$ is
in agreement with the cost of the brute force approach.

Similarly, the 2PCF costs ${\cal O}(N\theta_c^{-2})$ to compute.
This is because the computation of the 2PCF involves $M_2$ triplets
where
\begin{eqnarray}
  M_2 &\propto& \int_{r_{min}}^{r_{max}} \frac{dN}{dr} \frac{1}{\theta_c^2} dr \\
      &=& \frac{L^2}{\pi\theta_c^2} \int_{r_{min}}^{r_{max}} \frac{1}{r^3}dr\\
      &=& \frac{L^2}{2\pi\theta_c^2}
            \left(\frac{1}{r_{min}^2}-\frac{1}{r_{max}^2}\right) \\
      &=& \frac{L^2}{2\pi\theta_c^2}
            \left(\frac{\pi N}{L^2}-\frac{1}{L^2 \theta_c^2}\right) \\
      &=& \frac{1}{2\pi\theta_c^4}(\pi\theta_c^2 N - 1).
\end{eqnarray}
In the limit that $\theta_c^2 N\gg 1$,
\begin{equation}
  M_2 \propto \frac{N}{\theta_c^2},
\end{equation}
that is, the computational cost of the 2PCF scales as $N\theta_c^{-2}$.

  
The memory overhead of this new algorithm is ${\cal O}(N)$
since the tree is stored in a 1D array of length $2N-1$.

\subsection{\bf Tests of Computational Cost for 3PCF}
\label{subsec:test_speed}

As derived above, the computational cost of the raw 3PCF
(after tree building)
is expected to be ${\cal O}(N \theta_c^{-4} \ln(\theta_c^2 N))$
for $\theta_c^2 N \gg 1$.
Furthermore, for very large $\theta_c^2 N$, the scaling reduces to
${\cal O}(N \theta_c^{-4})$.

The speed of the algorithm is tested for $N$ up to $10^6$ galaxies
on CITA's 1.3GHz ItaniumII, Dell Poweredge 7250 computer.
The set of computations for speed testing utilizes
a randomly generated catalogue with the same attributes
as the mock galaxy catalogue used for accuracy comparison. 
The catalogue is a list of galaxies with
$x$, $y$ coordinates, $\gamma_1$, $\gamma_2$, $\kappa$ and noise.
Each of these attributes take random values between
their minimum and maximum.
While the coordinate components lie between $0$ and $54000$,
$\gamma_1$, $\gamma_2$ and $\kappa$ take values between $-1$ and $1$.
To ensure numerical stability, the values of the noise
is randomly selected in the range of $[1, 2)$.
The 3PCF is computed for triplets with separations between
$10$ and $\sqrt{2}\times 54000$ with binning at logarithmic
intervals of $2^{0.1}$.
Here, we test and analyze the computing time of tree construction
and that of the raw correlation function computation.

\begin{figure}
\vskip 3.3 truein
\includegraphics{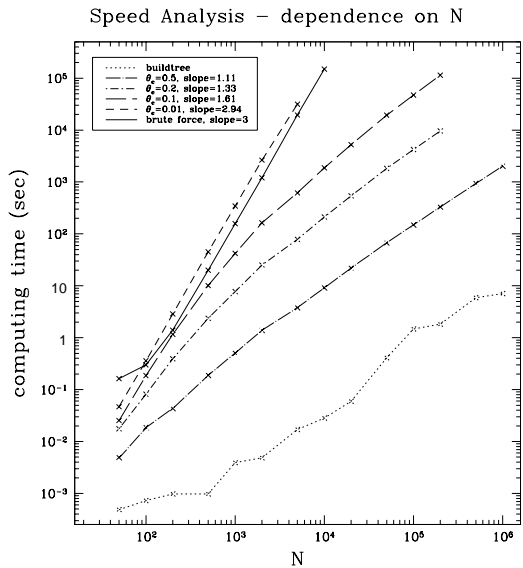}
\caption{A plot of the total raw 3PCF computing time for the
shear, $\kappa$ and the weight, against the number of particles.
While the raw correlation functions for $\kappa$ and for the
weight are scalar, the shear raw correlation function consists
of 8 components.
The different lines correspond to different $\theta_c$.
The slopes given in the plot are measured at a fixed
computing time of 1000 seconds.
One expects the slope to approach $1$ as
$\theta_c^2 N \rightarrow \infty$.
The plot also shows the average tree building time over the
different $\theta_c$'s.
}
\label{fig:speed_n}
\end{figure}
\begin{figure}
\vskip 3.3 truein
\includegraphics{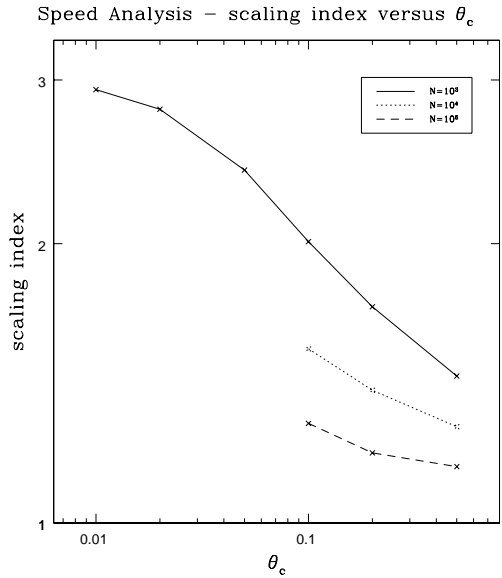}
\caption{A plot of the scaling index versus $\theta_c$
for different values of $N$.
Here, the scaling index is defined to be the log-log slope
of the computing time versus $N$.
On each curve plotted here, the scaling indices
are obtained by measuring the slopes in figure \ref{fig:speed_n}
at a fixed $N$.
However, curves for $\theta_c=0.02$ and $0.05$ are not shown
in figure \ref{fig:speed_n} to avoid crowdedness.
}
\label{fig:slope_t}
\end{figure}

Figure \ref{fig:speed_n} plots the total raw 3PCF computing time
for the shear, $\kappa$ and the weight as a function of the
number of galaxies, $N$.  
Since the cost of tree building is independent of $\theta_c$,
the plot also shows the average tree building time over
the different $\theta_c$'s.
In figure \ref{fig:speed_n}, we can see that the cost of the
raw 3PCF computation dominates in the range of $N$ considered.
This shows that the cost of tree building has a much smaller
prefactor than that of the raw correlation computation.
In the following discussions about speed testing, we focus
on the computation of raw correlation functions.
The curves for the raw correlation computations appear to
asymptotically approach the line with slope$=1$ as $N$ increases.
However, the curves as shown are not quite ${\cal O}(N)$ yet
since the convergence of $\ln(\theta_c^2 N)$ to a constant
as $\theta_c^2 N$ increases is slow.
Only when $\theta_c^2 N$ becomes quite a bit larger than those in
the plot, would the behaviour of the curves exhibit linear quality.
The slopes of the curves in figure \ref{fig:speed_n} are all
greater than one.  However, it is evident from the figure that
they decrease with increasing $\theta_c$ and $N$.
This relation is further illustrated in figure \ref{fig:slope_t}.

Now, let us take a look at the dependence of the slope in figure
\ref{fig:speed_n} on the value of $\theta_c$.
We call this slope the scaling index.
In figure \ref{fig:slope_t}, we plot the scaling index, which
is measured at fixed $N$, as a function of $\theta_c$.
We observe that the scaling index decreases monotonically
with $\theta_c$ for fixed $N$, and that it seems to
approach $1$ as $\theta_c \rightarrow \infty$.
On the other hand, the scaling index also decreases
monotonically with increasing $N$.
These behaviours are consistent with our theoretical estimate
that the dependence of the raw 3PCF computational cost
on N tends to a linear relation as $\theta_c^2 N$ increases.
We do not, in this work, study the detailed dependence of
these scaling indices on $\theta_c$ and $N$.

\begin{figure}
\vskip 3.3 truein
\includegraphics{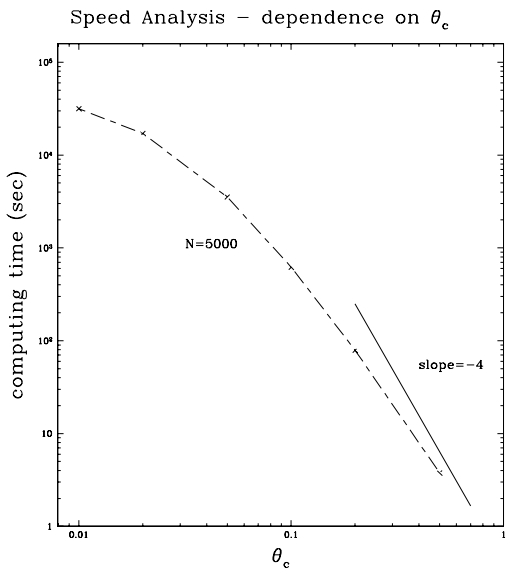}
\caption{A plot of the total raw 3PCF computing time for the
shear, $\kappa$ and the weight against $\theta_c$ for $N=5000$.
While the raw correlation functions for $\kappa$ and for the
weight are scalar, the shear raw correlation function consists
of 8 components.
A line with slope$=-4$, which is expected to be the limit for
large $\theta_c$, is also plotted for the purpose of comparison.
}
\label{fig:speed_t}
\end{figure}
In figure \ref{fig:speed_t}, we plot the total computing time
for the raw correlation functions of the shear, $\kappa$,
and the weight against the accuracy parameter $\theta_c$
while fixing $N$ to be $5000$.
The curve flattens out in the small $\theta_c$ limit.
This phenomenon is due to the extra factor of $\ln(\theta_c^2)$
in the computational cost as derived in section
\ref{subsec:estimate_speed}.
However, as $\theta_c$ increases, the curve appears to
asymptotically approach the line with slope$=-4$.
This is consistent with our expectation that the cost of
computing the raw 3PCF scales as $\theta_c^{-4}$
for very high values of $\theta_c^2 N$, that is,
when $\theta_c$ is sufficiently large for a fixed $N$.


\subsection{\bf Comparison of Computational Cost for 2PCF and 3PCF}
\label{subsec:comparison_speed}

Using our fast algorithm, the 2PCF in 2D costs
${\cal O}(N(\log N)^2 + N\theta_c^{-2})$ to compute
when $\theta_c^2 N \gg 1$.
On the other hand, for sufficiently large $\theta_c^2 N$,
the computing time of the 3PCF in 2D scales as
$N(\log N)^2 + N\theta_c^{-4}\ln(\theta_c^2 N)$.
The 3PCF scaling reduces to ${\cal O}(N\theta_c^{-4})$
as $\theta_c^2 N \rightarrow \infty$.
However, this convergence of the 3PCF computational cost to
a linear relation in $N$ is slow.

In comparison, using FFTs, the 2D two-point correlation function
can be computed at the cost of ${\cal O}(N \log N)$,
and the three-point correlation functions at
${\cal O}(N^2+N{\theta_c}^{-2} (\log N)^2)$.
The coefficient for FFTs can be very small, but the memory
overhead of ${\cal O}(N{\theta_c}^{-2}\log N)$ can be significant;
and in highly clustered regimes, there is a significant inefficiency
since the data must be gridded on the finest scale.

\citet{2001misk.conf...71M} described an algorithm to compute
the two-point function on a kd tree.  They reported a scaling of
${\cal O}(N^{3/2})$.  The additional cost appears to arise
because they loop over the correlation function bins and perform
a gather operation on pairs, while our algorithm performs a
scatter operation into the correlation function.

\section{Conclusion}
\label{sec:conclusion}

We have presented the framework to efficiently compute the $n$-point
correlation function.
This enables estimates of the full 3PCF in 2D
for $N$ as large as $10^6$.
Historically, the computation of the 3PCF has been prohibitively
expensive; the existing fast 2PCF algorithms often suffer loss of
generality due to geometric limitations.
In this work, we showed in detail how to compute the 2PCF and the
3PCF for a spin-2 field in the context of weak gravitational lensing.
In 2D and for sufficiently large $\theta_c^2 N$,
the 2PCF requires ${\cal O}(N (\log N)^2 + N\theta_c^{-2})$
to compute while the 3PCF computation can be
completed in ${\cal O}(N (\log N)^2 + N\theta_c^{-4}\ln(\theta_c^2 N))$
time where $\theta_c$ is
the critical open angle defined in section \ref{sec:theta}.
Recall that the ${\cal O}(N (\log N)^2)$ cost arises from the
tree construction, and the remaining cost is due to subdivisions
and node-to-node correlations.
Since the raw correlation function computation is the prime
consumer of computational effort in the range of consideration
(see figure \ref{fig:speed_n}), the total computational cost at a
fixed $\theta_c$ is dominated by ${\cal O}(N)$ and
${\cal O}(N\ln N)$ for the 2PCF and 3PCF respectively.
As $\theta_c^2 N \rightarrow \infty$, the cost of computing the
raw 3PCF tends to ${\cal O}(N)$.
On the other hand, when $\theta_c$ approaches zero,
one needs to descend all the way
down to the leaves to accumulate the desired statistics.
Therefore,
the computational cost limits to ${\cal O}(N^2)$ and ${\cal O}(N^3)$
respectively for the 2PCF and 3PCF, similar to the cost of
the brute force approach.
We also generalized the algorithm to compute correlation functions
of arbitrary order, with a discussion of its application to a scalar
field in higher dimensional space.

The technique involves the construction of a balanced
bisectional binary tree.
The worst case error at each node is minimized by dividing
perpendicular to the longest node extent.  Bisection guarantees
a tree of height no more than $\lceil 1+\log_2N \rceil$, and
a memory requirement no more than ${\cal O}(N)$.

At the same level of the tree hierarchy, all nodes are occupied by
approximately the same number of particles; and 
all leaves contain exactly one particle.
When walking the tree, we start from the 'root' node.
Whenever the open angle criterion is fulfilled, we accumulate
$n$-point statistics and terminate our pursuit along
that path; otherwise, we move down the tree until the criterion
is met by all $n$ nodes or until we hit the leaves.
Besides its speed, our tree method is favourable compared to
the computation of the full 3PCF using FFT
in that it is free of geometric restrictions and
that it can be easily extended to compute correlation functions of
arbitrary order.

For the upcoming lensing and CMB surveys such as the
Canada-France-Hawaii-Telescope-Legacy-Survey, this rapid algorithm
will allow an efficient analysis of the data in a tractable amount of
computational effort.

We would like to thank Robin Humble for suggesting the algorithm,
and Mike Jarvis for stimulating discussions.  We are grateful to
the referee for his valuable comments and useful suggestions.


\appendix
\section{APPENDIX: Scheme of mainstream subdivision for $\bmath{n}$-PCF}

The mainstream subdivision for the $n$-PCF is completed
by the following operations:
\begin{verbatim}
  j=0
  Do i=1,n-1
    If (nodes(i)=nodes(i+1)) then
      j=j+1
    Else
      If (j>0) then
        Do k=0,j+1
          snodes=nodes
          snodes(i-j:i-j+k-1)=nodes(i-j).sub1
          snodes(i-j+k:i)=nodes(i-j).sub2
          call SUBDIVIDE(snodes,n,c=0)
        Enddo
        j=0
      Endif
    Endif
  Enddo.
\end{verbatim}

\label{lastpage}
\end{document}